\documentclass{IEEEtran}
\usepackage{cite}
\usepackage{amsmath,amssymb,amsfonts}
\usepackage{algorithmic}
\usepackage{graphicx}
\usepackage{textcomp}
\usepackage{verbatim}
\usepackage[colorlinks=true, linkcolor=red, citecolor=blue, urlcolor=green]{hyperref}
\usepackage[numbers,sort&compress]{natbib}
\newcommand{\raisedchi}{\raisebox{\depth}{\(\chi\)}}

\usepackage{caption}
\usepackage{subcaption}
\usepackage{cleveref} 
\crefname{equation}{}{}

\def\BibTeX{{\rm B\kern-.05em{\sc i\kern-.025em b}\kern-.08em
    T\kern-.1667em\lower.7ex\hbox{E}\kern-.125emX}}
\begin{document}
\title{A 3D Formulation of the Extended Phaseless Rytov Approximation}
\author{Wanqin Ma, \IEEEmembership{Graduate Student member, IEEE}, Zan Li, \IEEEmembership{Member, IEEE}, Amartansh Dubey, \IEEEmembership{Member, IEEE}, Alikhan Umirbayev, \IEEEmembership{Graduate Student member, IEEE}, Yijun Chen, \IEEEmembership{Graduate Student member, IEEE}, Junhui Rao, \IEEEmembership{Member, IEEE}, and Ross Murch, \IEEEmembership{Fellow, IEEE}
\thanks{Manuscript received June 5, 2026. This work was supported by the Hong Kong Research Council Area of Excellence Grant  AoE/E-601/22-R. \textit{(Corresponding author: Amartansh Dubey)}}
\thanks{Wanqin Ma, Zan Li, Alikhan Umirbayev, Yijun Chen, Junhui Rao, are with the Department of Electronic and Computer Engineering, Hong Kong University of Science and Technology (HKUST), Hong Kong (e-mail: wmaag@connect.ust.hk) }
\thanks{Amartansh Dubey is with the department of Electrical Engineering, Indian Institute of Technology, Delhi, India (e-mail: amardubey@ee.iitd.ac.in)}
\thanks{Ross Murch is with the Department of Electronic and Computer Engineering, Institute of Advanced Study, Hong Kong University of Science and Technology (HKUST), Hong Kong (e-mail: eermurch@ust.hk)}}

\maketitle

\begin{abstract}
The extended Phaseless Rytov Approximation (xPRA) is a recently proposed device-free RF imaging technique that provides high-resolution reconstructions of the imaging region using only phaseless measurements, such as received signal strength (RSS). Because of its phaseless formulation, it can be implemented straightforwardly using existing wireless communication infrastructure. It also outperforms well-known device-free phaseless RF imaging methods such as Radio Tomographic Imaging (RTI). The linear phaseless formulation used in xPRA (and RTI) makes these methods potentially useful for integrated sensing and communication (ISAC) systems in next generation wireless networks since they do not require wide bandwidths. However, so far, both xPRA and RTI have primarily been formulated in two dimensions (2D). This paper introduces a 3D extension of xPRA, which we call the extended three-dimensional phaseless Rytov approximation (x3DPRA). The novelty of our approach is that it preserves the straightforward implementation advantages of RTI and xPRA while enabling volumetric (3D) imaging. Simulation results show that x3DPRA provides good estimates of location and shape and can also reconstruct object material attenuation. We present the 3D formulation, validate it with a 2D model comparison, and report simulation results demonstrating its performance.
\end{abstract}


\section{Introduction}
\label{sec: introduction} 
\IEEEPARstart{T}{he} plan for the next generation of wireless networks extends beyond communication, aiming to combine communication with sensing to perceive the physical environment while also providing communication. Integrated Sensing and Communication (ISAC) is built on this concept, in which radio frequency (RF) signals are used for both data transmission and environmental perception~\cite{ISAC,AA,BB,CC}. Specifically, RF sensing analyzes the propagation and reflection of radio signals to capture environmental information. This technology has various applications, including device-free object detection, tracking, localization, and RF imaging~\cite{RF_sensing}. Compared to other vision-based approaches, such as optical cameras and LiDAR, RF sensing is more robust in poor lighting, privacy-sensitive areas, and at long distances. Furthermore, its ability to travel through various materials enables non-intrusive imaging through walls~\cite{Camara, RF_2}.

Within RF sensing, the branch that focuses on detecting and localizing targets without carrying any active transceivers is known as device-free sensing. Device-free sensing is more convenient than device-based methods, since target objects can be tracked without any device attached.  Several studies investigate device-free imaging to estimate the shape, size, and properties of target objects, including indoor imaging~\cite{Indoor_1, Indoor_2, Indoor_3, Indoor_4}, compressive sensing~\cite{Compress_1, Compress_2, Compress_3, Compress_4}, and statistical models~\cite{Stat_1, Stat_2, Stat_3, Stat_4}. Indoor imaging has several important applications, including security systems, smart homes, indoor navigation, and health monitoring. 

In this work, we focus on imaging that uses only phaseless, or equivalently, received signal strength (RSS) data for device-free sensing. Using phaseless data greatly simplifies the system by eliminating the need for coherent synchronization between sensor nodes. Furthermore, these techniques usually do not require calibration because their linear formulations enable temporal subtraction~\cite{Amar_TGRS}. These techniques not only alleviate the intricate calibration required in phase-based imaging systems but also can enhance the imaging quality by removing background scattering. For these reasons, the sensor nodes can be based on existing infrastructure such as Wi-Fi, cellular, or Bluetooth. Another key advantage of the proposed techniques is that they are single-frequency, with resolution proportional to the propagation wavelength. This is in contrast to radar-based techniques~\cite{Radar_1, Radar_2, Radar_3, Radar_4, Radar_5, 3D_Radar_Car}. While radar methods offer accurate reconstruction, their range resolution is constrained by bandwidth. For example, discriminating features smaller than about 15 cm would require a bandwidth exceeding 1 GHz, which is unrealistic for most commercial operations. This restriction cannot meet the requirements for imaging with limited bandwidth. 

One widely investigated phaseless device-free imaging technique in wireless systems is radio tomographic imaging (RTI)~\cite{RTI}. RTI reconstructs an indoor environment by analyzing changes in Wi-Fi signal strength. It is configured by placing sensors at the same height around a 2D domain of interest (DOI) to form a wireless network. When an object, such as a person, moves through this network, it attenuates the radio signals passing through it. RTI uses the measured RSS drop across multiple links to reconstruct a 2D attenuation map that determines the shape and location of objects. This method simplifies the complex radio environment by focusing on the direct line-of-sight (LOS) attenuation and modeling the scene on a single 2D plane. RTI is largely based on an empirical formulation, and while being very straightforward, it provides good reconstructions in a surprisingly large number of indoor environments.
RTI has applications in areas such as human-path tracking, through-wall imaging, and hybrid systems with reconfigurable intelligent surfaces~\cite{RTIA_1,RTIA_2,RTIA_3, RTIA_4}. Based on RTI, ~\cite{Amar_xPRA, Amar_xRTI} developed a 2D phaseless extended Rytov approximation (xPRA) model by introducing more formal electromagnetic wave propagation phenomena. It enables the 2D reconstruction not only of location and shape but also of the permittivity of targets. This represents a significant step forward, as it provides more accurate and physically meaningful results. However, all these methods are primarily limited to 2D scenarios. Their performance and application in 3D environments have not been well investigated. 

A few works have attempted to extend RTI into three dimensions. One approach has been to model multi-height LOS with voxels in 3D space. For data collection, these works utilize a network of Wi-Fi sensors positioned at various heights around the 3D DOI. It has been applied to road surveillance~\cite{AFIT_Road, NYTU_Road}, as well as indoor human path tracking and localization~\cite{DanellaThesis}. However, these methods focus on localization, which tracks the centroid of a moving target. Another approach is to apply RTI across multiple height layers by scanning with additional devices, then stacking the 2D results to form a 3D reconstruction. Furthermore, using drones and transceiver slide rails carrying Wi-Fi sensors, one can estimate a general 3D shape of targets by scanning them vertically, layer by layer, around the DOI~\cite{3D_WIFI_Drone, Rice}. Because the results from these works at each height layer are not well integrated, the composite 3D reconstruction often appears disjointed and lacks vertical smoothness. Most importantly, none of the existing RTI methods can distinguish target materials by combining electromagnetic wave propagation. It limits the application of RTI in the real world because, without material classification, the system cannot accurately identify an object, thereby reducing its value for applications such as smart buildings. Thus, it is important to extend propagation-based RTI methods, such as xPRA, to 3D. 

In this work, we develop a 3D framework utilizing xPRA for advanced RF imaging, enabling full volumetric reconstruction of environments and target properties. The key objective of this work is to provide 3D reconstructions of the attenuation parameter. Our contributions include: 
\begin{enumerate}
    \item We formulate a 3D extension of xPRA, which we denote as x3DPRA. It maintains the objectives of RTI and xPRA, which are to provide a straightforward technique that requires no calibration and allows background subtraction to remove clutter. 
    \item Develop numerical techniques to obtain reconstructions using the x3DPRA formulation using measurement data. A 3D regularization technique is utilized in the optimization required to obtain the reconstructions. 
    \item Provide simulation results in a realistic imaging scenario. We consider a 3D DOI with a volume of $0.9 \times 0.9 \times 0.3 \text{m}^3$ and reconstruct the attenuation parameter of the object in the DOI.
\end{enumerate}
The remainder of this paper is organized as follows. In Section~\ref{sec: PF}, we describe the imaging problem and necessary notation. Section~\ref{sec: M} formulates x3DPRA by extending xPRA from 2D. Section~\ref{sec: RR} describes the numerical solution for our x3DPRA formulation. Section~\ref{sec: R} presents and analyzes the current simulation results. Finally, Section~\ref{sec: C} concludes the paper by summarizing the key findings and contributions and discussing future directions based on x3DPRA.

\section{Problem Formulation}
\label{sec: PF}
\subsection{Problem Setup}
\begin{figure}[htbp]
	\centering
	\begin{subfigure}[b]{\linewidth}
    \centering
	    \includegraphics[width=0.85\linewidth]{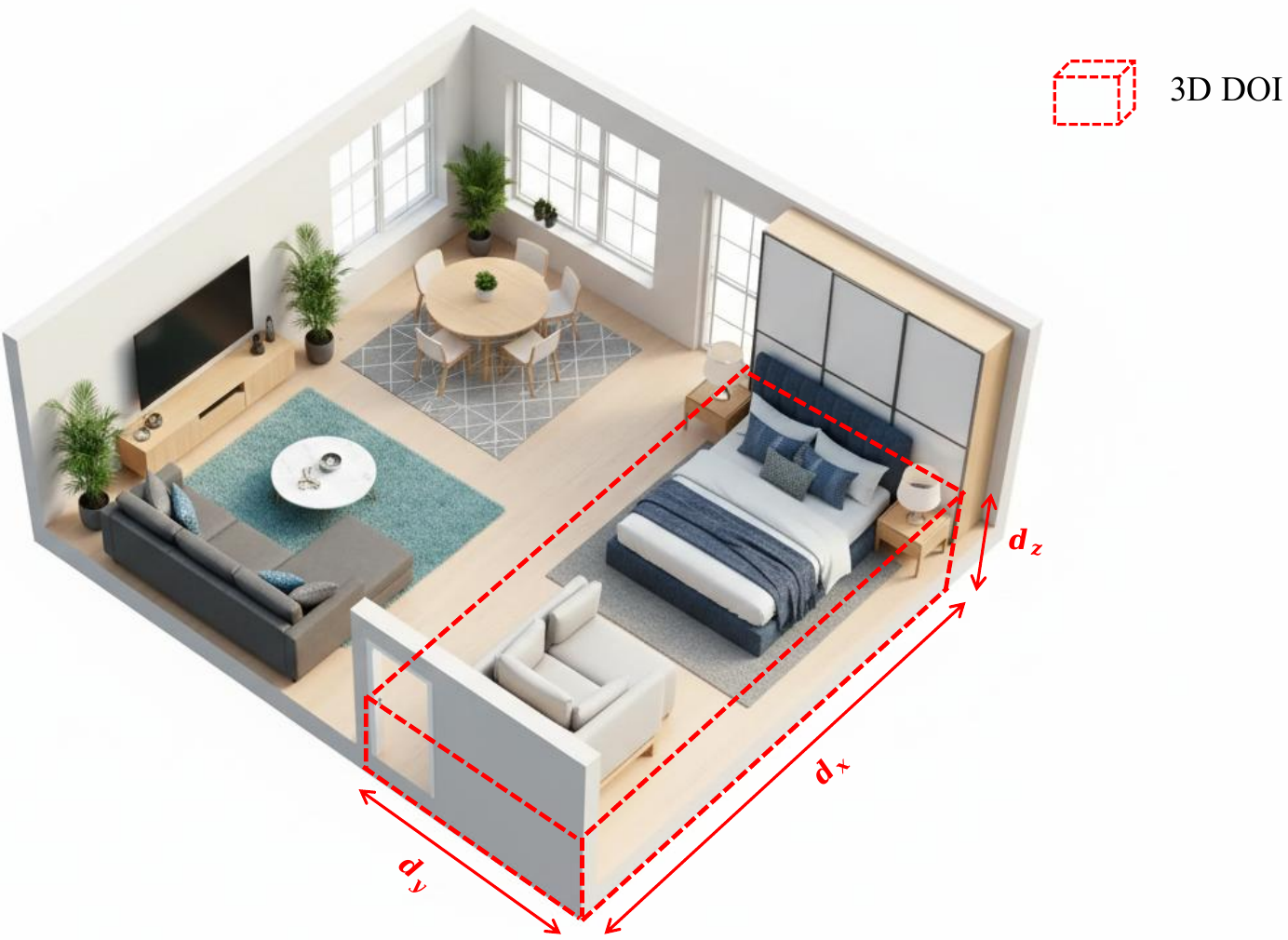}
        \caption{An example of a 3D indoor scenario showing the DOI in the red box.}
     \end{subfigure}
     \vfill
     \begin{subfigure}[b]{\linewidth}
     \centering
	    \includegraphics[width=\linewidth]{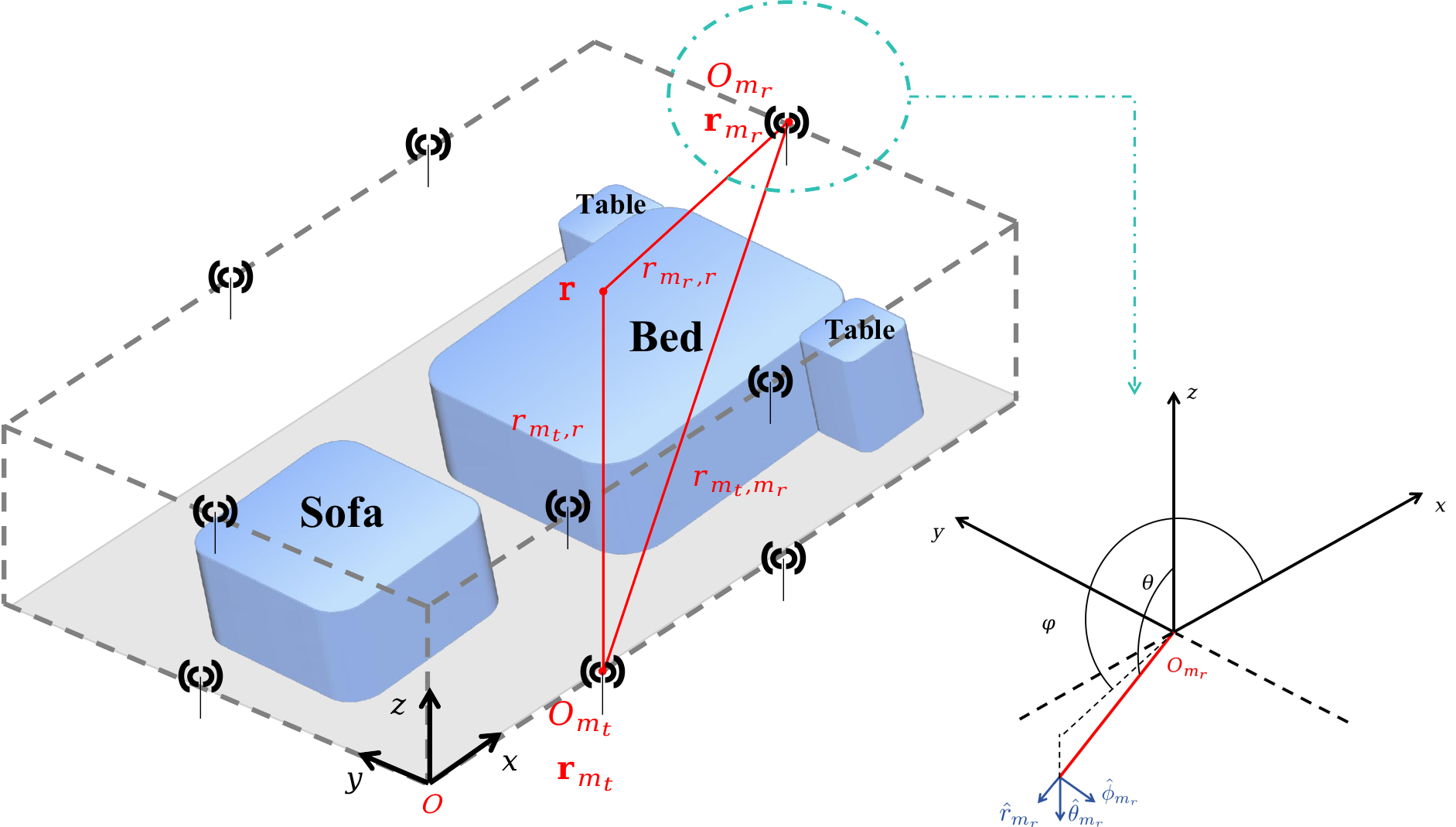}
        \caption{The 3D DOI where transceiver nodes are shown on the boundary $\mathcal{B}$ with the common origin $O$ of the rectangular coordinate system shown. It is also convenient to define origins relative to each transmitter and receiver, denoted by $O_{m_t}$ and $O_{m_r}$, respectively. For clarity, an example of a spherical coordinate system with origin $O_{m_r}$ is written as $\hat{r}_{m_r}$, $\hat{\theta}_{m_r}$,$\hat{\phi}_{m_r}$, and is shown. }
        \label{fig: 3D_DOI_b}
     \end{subfigure}
	\caption{This image shows a typical imaging scenario with a real-life indoor example in (a) and details of DOI geometry in (b).}
	\label{fig: 3D_DOI}
\end{figure}

Consider an example DOI consisting of the scenario inside a building as shown in Fig~\ref{fig: 3D_DOI}. Identical 2.4 GHz Wi-Fi transceivers are set up around the boundary of the DOI, for collecting measurement data as shown in Fig~\ref{fig: 3D_DOI_b}. There are $L$ transceivers in total, which transmit and receive signals to one another. However, a single transceiver cannot receive and transmit simultaneously. Each pair of transceivers creates a link, and the number of such unique links is $M = L(L-1)/2$ (excluding the reciprocal links). To specify the transceiver locations, we use coordinates relative to a common origin $O$ as shown in Fig~\ref{fig: 3D_DOI_b}. A point inside the DOI is shown as $\mathbf{r}$. The set of points $\mathbf{r}$ inside the DOI is defined as $\mathcal{D} \subset \mathbb{R}^3$ while those on the boundary are defined by $\mathcal{B} \subset \mathbb{R}^3$. The $z$-axis coordinate is oriented in the vertical direction, and all transceivers are assumed to have vertically polarized antennas. 

We denote the transceiver node operating as a transmitter at $\mathbf{r}_{m_t}$ as $m_t$, and the node operating as a receiver at $\mathbf{r}_{m_r}$  as $m_r$. We refer to the resulting link as $l_{m_t,m_r}$ and the distance between the two nodes as $r_{m_t,m_r}$. 
In the following, we will also refer to the transceiver node directly as a transmitter or a receiver when operating in the corresponding mode.  
Coordinates relative to each transmitter and receiver are also introduced. The origins of these coordinate systems with respect to $O$ are located at $\mathbf{r}_{m_t}$ and $\mathbf{r}_{m_r}$ and are denoted as $O_{m_t}$ and $O_{m_r}$ respectively. The rectangular coordinate systems for these are all affine, so their $z$-axes are also vertical. Spherical coordinates with respect to these origins are also useful: the basis vectors referred to $O_{m_t}$ are $\hat{r}_{m_t}$, $\hat{\theta}_{m_t}$,$\hat{\phi}_{m_t}$; while for $O_{m_r}$, the basis vectors are $\hat{r}_{m_r}$, $\hat{\theta}_{m_r}$,$\hat{\phi}_{m_r}$ as shown in Fig~\ref{fig: 3D_DOI_b}. The link $l_{m_t,m_r}$ with respect to the transmitter origin $O_{m_t}$ has its location written as $ (\mathbf{r}_{m_t, m_r}, \theta_{m_t,m_r},\phi_{m_t,m_r} )$ in spherical coordinates where the first subscript refers to the origin utilized,  $O_{m_t}$. The location of an arbitrary point $\mathbf{r}$ with respect to origin $O_{m_t}$ is written as $ (\mathbf{r}_{m_t, r}, \theta_{m_t,r},\phi_{m_t,r} )$ similarly. Vectors $\mathbf{r}_{m_t,m_r}$ and $\mathbf{r}_{m_r,m_t}$ have the same length $r_{m_t,m_r}$ but different directions. Similarly the length of $\mathbf{r}_{m_t,r}$ and $\mathbf{r}_{m_r, r}$ is $r_{m_t,r}$ and $r_{m_r,r}$, respectively.

We describe the material property of interest inside the 3D DOI (i.e points $\mathbf{r} \in \mathcal{D}$),  by relative permittivity $\epsilon_r (\mathbf{r}) = \epsilon_R(\mathbf{r}) + j\epsilon_I(\mathbf{r})$. The time variations of objects are taken as negligible over the time it takes for a radio wave to propagate across the DOI. Furthermore, as most of the objects around us are low-loss~\cite{Amar_xPRA}, we utilize the attenuation parameter~\cite{Att_1, Att_2} as the material property we reconstruct. The definition of the attenuation parameter under the low-loss condition can be written as~\cite{Att_1, Att_2}.
\begin{equation}
\label{eq: att}
    \alpha(\mathbf{r}) = \frac{2\pi \epsilon_I(\mathbf{r})}{\lambda_0 \sqrt{\epsilon_R(\mathbf{r})}} = \frac{2\pi\delta\sqrt{\epsilon_R(\mathbf{r})}}{\lambda_0}
\end{equation}
where $\lambda_0 = 0.125$ m is the wavelength for 2.4 GHz Wi-Fi signals, $\delta = \frac{\epsilon_I(\mathbf{r})}{\epsilon_R(\mathbf{r})}$ is the loss-tangent. In this way, $\alpha(\mathbf{r} )$ can be used to describe the attenuation profile of the DOI at any time instant $t$.

For the $l_{m_t,m_r}$th link, the free space incident field at $m_r$ due to a transmitter at $m_t$ is $\mathbf{E}^i_{m_t}(\mathbf{r}_{m_r})$ while $\mathbf{E}^i_{m_t}(\mathbf{r})$ is the incident field at $\mathbf{r}$.  $\mathbf{E}_{m_t}(\mathbf{r}_{m_r})$ and $\mathbf{E}_{m_t}(\mathbf{r})$ refers to the total electric field at $m_r$ and $\mathbf{r}$ respectively.

\subsection{Transceiver Antenna Model}

The transceivers are fitted with vertically polarized antennas that are circularly symmetric along their $z$-axes, such as dipoles and monopoles. The radiated field by the $m_t$ transmitter at point $\mathbf{r}$ is written
 \begin{equation}
 	\label{eq: pattern}
 	\mathbf{E}^i_{m_t}(\mathbf{r}) =  P(\theta_{m_t,r}) g(\mathbf{r}-\mathbf{r}_{m_t}) \hat{\theta}_{m_t}
 \end{equation}
where $g(\cdot)$ is the free space Green's function and $P(\theta_{m_t,r})$ is the radiation pattern of the antenna.  Since the antennas are vertically polarized and circularly symmetric, there is no variation in $\phi_{m_t}$ and no electric field component along $\hat{\phi}_{m_t}$. The superscript $i$ is used to denote the incident field and indicates that the radiated fields from the transceivers are incident fields for this configuration.

The voltage at receiver $m_r$ due to the field caused by the incoming wave can also be obtained straightforwardly~\cite{Collin1985Antennas,Amar_TVT}. For example, the voltage received due to the incident field from transmitter $m_t$ is 
\begin{equation}
	\label{eq: Vol}
	V_{m_t}\left(\mathbf{r}_{m_r}\right)=\mathbf{E}_{m_t}^i(\mathbf{r}_{m_r}) \cdot h(\theta_{m_r,m_t})
\end{equation}
where $\theta_{m_r,m_t}$ is the angle of the incoming wave, $h\left(\theta_{m_r,m_t}\right)$ is the antenna height, and for a standard half-wavelength dipole can be written as
\begin{equation}
	\label{eq: h}
	h\left(\theta_{m_r,m_t}\right)=\frac{\lambda_0}{j} \sqrt{\frac{R}{\pi \zeta_o}}\sin(\theta_{m_r,m_t})
\end{equation}
where $R$ is the antenna input resistance, and $\zeta_o$ is the impedance of free space.

\subsection{Measurement Dataset}

Since measurements can only be taken at the DOI boundary $\mathcal{B}$ we can only collect  $\mathbf{E}_{m_t}(\mathbf{r}_{m_r})$ at points $\mathbf{r}_{m_r}$. Therefore the measurement data set consists of $\mathbf{E}_{m_t}(\mathbf{r}_{m_r})$ across all transceiver pairs $m_t$ and $m_r$ providing $M=L(L-1)/2$ unique measurements. We denote this set of measurements as $\mathcal{M}$. Using $\mathcal{M}$ we need to estimate $\alpha(\mathbf{r})$ over  $\mathcal{D}$. 

We also define measurement sets that are subsets of $\mathcal{M}$. These include 2D cross-sectional measurements, defined as those in which the transmitter-receiver pairs all lie on a 2D plane. If all the pairs are at the same height, then it is a horizontal plane. However, the plane can also be inclined. This measurement set corresponds to previous 2D imaging scenarios, and we refer to such datasets as $\mathcal{M}_{2D}$.  If there are measurements $\mathcal{M}_{2D}$ at two or more 2D planes, we can reconstruct multiple cross sections. We refer to this as 2.5D with dataset $\mathcal{M}_{2.5D}$. The key here is that the measurement links between planes are not included in $\mathcal{M}_{2.5D}$. If measurements for all transmitter-receiver pairs are acquired, we have a full 3D configuration, which we refer to as 3D imaging with dataset $\mathcal{M}_{3D}$ or, more simply, $\mathcal{M}$.

The major difference between the 3D and 2.5D configurations is that in 3D, the measurement data is supplemented with measurements between planes, thereby increasing the amount of available data. Therefore, the 3D configuration should provide reconstructions that are better than those in 2.5D, since more data is acquired. 

\subsection{Problem Statement}
\label{sub: Problem}
This work aims to reconstruct $\alpha(\mathbf{r})$ over $\mathbf{r} \in \mathcal{D}$ using phaseless measurements of the received voltages across the dataset $\mathcal{M}_{3D}$ which have been obtained using measurements $\mathbf{E}_{m_t}(\mathbf{r}_{m_r})$ where $\mathbf{E}^i_{m_t}(\mathbf{r}_{m_r})$ is known. The phaseless measurements can be taken as the absolute value of the received voltages $|V_{m_t}\left(\mathbf{r}_{m_r}\right)|$ or as RSS in terms of received power $P_{m_t}\left(\mathbf{r}_{m_r}\right)$ (defined later).

\section{Methodology}
\label{sec: M}

\subsection{3D Phaseless Extended Rytov Approximation}
\label{subsec: ass}

The governing equation for the electric field at point $\mathbf{r} \in \mathcal{D}$ due to a transmitter at $m_t$, is given by by~\cite{Chen2017Computational} 

\begin{equation}
\label{eq: Chen1}
\nabla \times \nabla \times \mathbf{E}_{m_t}(\mathbf{r})-k_0^2 \mathbf{E}_{m_t}(\mathbf{r})= k_0^2 (\epsilon_r(\mathbf{r})-1)\mathbf{E}_{m_t}(\mathbf{r})
\end{equation}
where $k_0$ is the wavenumber in free space. In volume source integral (VSI) form, (\ref{eq: Chen1}) can be written as 
\begin{equation}
	\label{eq: Chen2}
	\begin{aligned}
\mathbf{E}_{m_t}(\mathbf{r}_{m_r})&=\mathbf{E}^{i}_{m_t}(\mathbf{r}_{m_r})  \\ & + k_0^2  \iiint (\epsilon_r(\mathbf{r}) - 1) \overline{\overline{G}}\left(\mathbf{r}_{m_r}, \mathbf{r}\right) \cdot \mathbf{E}_{m_t}\left(\mathbf{r}\right) d \mathbf{r}
\end{aligned}
\end{equation}
where $\overline{\overline{G}}\left(\cdot, \cdot\right)$ is the dyadic Green's function. 

The 3D model (\ref{eq: Chen2}) includes intricate polarization and scattering effects, and performing full inversion requires detailed measurements, including calibration, as well as an intensive inversion process~\cite{3D_ISP1,3D_ISP2,3D_ISP3}. To maintain the straightforward approaches used in RTI and xPRA, our objective is to extend these techniques approximately to 3D without the intricate methods required to fully solve (\ref{eq: Chen2}). 

In 2D, the transverse magnetic (TM) formulation is utilized in xPRA under the assumption that objects in the DOI can be approximated by infinitely long cylinders with their axis oriented along the $z$-axis. With this assumption, object cross sections can be reconstructed. The quality of the reconstruction depends on how well the TM approximation holds. In configurations where objects of interest exhibit vertical variations over lengths greater than one wavelength, the assumptions are valid. As such, reconstructions have been shown to be excellent in many scenarios, as shown for both simulation and experimental results~\cite{Amar_xPRA,Amar_xRTI,RTIA_1, RTIA_4}. One reason the RTI and xRA approaches perform well is background subtraction that leverages the sparse indoor environment, which consists mainly of free space, as discussed later. 

To extend xPRA to 3D while maintaining the straightforward approach of 2D xPRA, we need to use related assumptions to side-step the intricate 3D polarization and scattering effects. The first step in such an approach is to use vertically polarized antennas, as described in our measurement configuration in the previous section. These only have a $\hat{\theta}_{m_t}$ radiation component, which simplifies the handling of polarization. With this configuration, we can define a plane passing through both $m_t$ and $m_r$ with the plane's normal parallel to $\hat{\theta}_{m_t,m_r}$. We denote this plane as $\mathcal{P}_{m_t,m_r} \subset \mathbb{R}^2$. Between the two transceiver pairs $m_t$ and $m_r$, waves are TM on this plane, $\mathcal{P}_{m_t,m_r}$, when there is no scattering. When objects are present, scattering will occur, and to handle this, we can extend the 2D assumptions described in the previous paragraph. In particular, we can approximate objects that intersect the plane, $\mathcal{P}_{m_t,m_r}$, as infinite cylinders with their axes oriented along the plane's normal. The validity of this assumption is similar to that of the 2D case. In configurations where objects of interest vary in length by more than one wavelength, it is approximately valid. Furthermore, taking transceiver pairs on planes $\mathcal{P}_{m_t,m_r}$ that are not significantly inclined from the horizontal will also aid this approximation.

With the observation that the electric field can be taken as TM in planes $\mathcal{P}_{m_t,m_r}$ we can write $\mathbf{E}_{m_t}(\mathbf{r})=E_{m_t}(\mathbf{r})\hat{\theta}_{m_t,m_r}$, where $E_{m_t}(\mathbf{r})$ is a scalar. Similarly we also write $\mathbf{E}^{i}_{m_t}(\mathbf{r}_{m_r})$ as the scalar $E^i_{m_t}(\mathbf{r}_{m_r})$, $\mathbf{E}_{m_t}(\mathbf{r}_{m_r})$ as scalar $E_{m_t}(\mathbf{r}_{m_r})$ and $\mathbf{E}^{i}_{m_t}(\mathbf{r})$ as the scalar $E^i_{m_t}(\mathbf{r})$. 

Using the scalar forms we can re-write~\cref{eq: Chen1} and~\cref{eq: Chen2} in scalar form for when $\mathbf{r} \in \mathcal{P}_{m_t,m_r}  $ as
\begin{equation}
\label{eq: Amar0}
\nabla^2 {E}_{m_t}(\mathbf{r})+k_0^2 {E}(\mathbf{r})= k_0^2 (\epsilon_r(\mathbf{r})-1){E}_{m_t}(\mathbf{r})
\end{equation}
\begin{equation}
	\label{eq: Amar1}
	\begin{aligned}
		E_{m_t}(\mathbf{r}_{m_r}) &= E^i_{m_t}(\mathbf{r}_{m_r}) +\\& k_0^2 \int_{\mathcal{P}} g(\mathbf{r}_{m_r}, \mathbf{r}) (\epsilon_r(\mathbf{r})-1) E_{m_t}(\mathbf{r}) d\mathbf{r}^2
	\end{aligned}
\end{equation}
respectively. The VSI, \cref{eq: Amar1}, is also known as the Lipmann-Schwinger equation \cite{Mittra1998, Chen2017Computational}. 
 
To perform reconstruction we need to solve (\ref{eq: Amar1}) for $\epsilon_r(\mathbf{r})$. Since (\ref{eq: Amar1}) is an extended 3D version of an underlying 2D form, we can also extend our previous 2D reconstruction approaches to 3D in a similar way. To perform this, we use our extended Rytov approximation for low-loss media \cite{Amar_xPRA}. In previous 2D work~\cite{Amar_xPRA}, the Rytov approximation was extended and formulated in 2D as xPRA. The simulation and experimental results demonstrated that high-quality reconstructions could be obtained over a wide range of permittivity values, making this approach suitable for indoor RF imaging. 

In the Rytov approach the total field $E_{m_t}(\mathbf{r})$ is normalized by the incident field $E^i_{m_t}(\mathbf{r})$ so it can be expressed as a complex phase $\phi_{m_t}(\mathbf{r})$,
\begin{equation}
	\label{Eq_rytTfield}
	\begin{aligned}
		\frac{E_{m_t}(\mathbf{r})}{E^i_{m_t}(\mathbf{r})} &= e^{\phi_{m_t}(\mathbf{r}) }
	\end{aligned}
\end{equation}
We can think of $\phi_{m_t}(\mathbf{r})$ as representing the phase and log amplitude deviations from the incident field due to scattering.


Substituting (\ref{Eq_rytTfield}) into (\ref{eq: Amar1}) provides the Riccati equation \cite{wu2003wave},
\begin{subequations}
	\label{Eq_rytdiffeq3}
	\begin{align}
		(\nabla^2 +  k_0^2) (E^i_{m_t}(\mathbf{r}) \phi_{m_t}(\mathbf{r})) = -k_0^2 \raisedchi_{\text{RI}} (\mathbf{r}) E^i_{m_t}(\mathbf{r}) \\
		\raisedchi_{\text{RI}}  (\mathbf{r}) = \epsilon_r(\mathbf{r})-1 + \frac{\nabla \phi_{m_t}(\mathbf{r}) \cdot \nabla \phi_{m_t}(\mathbf{r})}{k_0^2}	\end{align}
\end{subequations}
where $\raisedchi_{\text{RI}}$ is referred to as the contrast function. Equation (\ref{Eq_rytdiffeq3}) can be written in integral form (denoted here as the Rytov integral (RI)) for the total field,
\begin{equation}
	\label{Eq_rytov2}
    \begin{aligned}
        E_{m_t}(\mathbf{r}_{m_r})  &= E^i_{m_t}(\mathbf{r}_{m_r})\cdot \\& \exp\bigg (  \frac{k_0^2}{E^i_{m_t}(\mathbf{r}_{m_r})} \int_{\mathcal{P}}  g(\mathbf{r}_{m_r}, \mathbf{r})  \raisedchi_{\text{RI}}(\mathbf{r}) E^i_{m_t}(\mathbf{r}) d\mathbf{r}^2\bigg)
    \end{aligned}
\end{equation}

The term $\nabla \phi_{m_t} \cdot \nabla \phi_{m_t}$ in $\raisedchi_{\text{RI}}$  is then neglected under a weak scattering assumption to arrive at the Rytov approximation. However, an extension of the 2D Rytov approximation(xRA) has been developed and can be applied here. This xRA relies on approximating $\phi_{m_t}$ using rays in low-loss media, which allows the term $\nabla \phi_{m_t} \cdot \nabla \phi_{m_t}$ to be partly modeled rather than neglected. This has been shown to substantially enhance reconstruction quality~\cite{Amar_xPRA,Amar_xRTI}. Using xRA, the contrast can be written as 
\begin{equation}
\label{Eq_xRA}
	\begin{aligned}
		\raisedchi_{\text{RI}}(\mathbf{r})  =2 (\sqrt{\epsilon_R(\mathbf{r})} -1)	+ j  \frac{\epsilon_I(\mathbf{r})}{\sqrt{\epsilon_R(\mathbf{r})}}
	\end{aligned}
\end{equation}
The importance of this form is that the second term on the right-hand side can be reconstructed very well over a wide range of permittivities. The imaginary part of this term is directly proportional to our attenuation parameter (\ref{eq: att}) through wavenumber $k_0$.

\subsection{Phaseless Voltage and Power Sensing}
In practice, we can only sense the voltage induced by the electric field incident on the receiving antenna; therefore, (\ref{Eq_rytov2}) needs further development. In addition to maintaining the straightforward nature of RTI and xPRA, these voltages should be in magnitude or phaseless form so that RSS or received power is required only.

To modify~\cref{Eq_rytov2} so that it is in terms of voltage, we need to refer to~\cref{eq: Vol} and use the concept of antenna height to obtain the voltage at the antenna terminals in terms of the incoming electric field. If the antennas were completely omnidirectional, no modifications of the formula~\cref{Eq_rytov2} would be required, as the received voltage would be directly related to the total field at the receiver by a straightforward proportionality constant. However, because the antennas have radiation patterns that vary with angle, we need to account for this angular dependence. To perform this, we can substitute~\cref{eq: Vol} into~\cref{Eq_rytov2}  so that the dependence on antenna height is explicitly included in~\cref{Eq_rytov2}. Using this the received voltage $V(\mathbf{r}_{m_r})$ can be written as
\begin{equation}
	\label{eq: vol_RI}
	\begin{aligned}
		V(\mathbf{r}_{m_r}) = &E^i_{m_t}(\mathbf{r}_{m_r}) h(\theta_{m_r,m_t})\exp\left( \frac{k_0^2}{E^i_{m_t}(\mathbf{r}_{m_r}) h(\theta_{m_r,m_t})} \right.\cdot \\
		&\left. \int_{{\cal P}} g\left( \mathbf{r}_{m_r}, \mathbf{r} \right) \chi_{\text{RI}}\left( \mathbf{r} \right) E^i_{m_t} \left( \mathbf{r} \right) h(\theta_{{m_r},r}) d\mathbf{r}^2 \right)
	\end{aligned}
\end{equation}

The voltage formulation (\ref{eq: vol_RI}) is useful for practical implementations; however, it still requires phase measurement. An even more useful approach is to use phaseless sensing, where we measure only power or RSS. This can be obtained by multiplying (\ref{eq: vol_RI}) by its conjugate. If we also take $\log_{10}$ both sides, we can obtain the result in terms of the total received power $P_{m_t}(\mathbf{r}_{m_r})$ and incident power $P^i_{m_t}(\mathbf{r}_{m_r})$ (in dB), 
\begin{equation}
\label{Eq_RytovApp}
    \begin{aligned}
        P_{m_t}(\mathbf{r}_{m_r})  = &P^i_{m_t}(\mathbf{r}_{m_r})+ C_0  \cdot \operatorname{Re}\left( \frac{k_0^2}{E^i_{m_t}(\mathbf{r}_{m_r}) h(\theta_{m_r,m_t})} \right.\cdot \\
		&\left. \int_{{\cal P}} g\left( \mathbf{r}_{m_r}, \mathbf{r} \right) \chi_{\text{RI}}\left( \mathbf{r} \right) E^i_{m_t} \left( \mathbf{r} \right) h(\theta_{{m_r},r}) d\mathbf{r}^2 \right)
    \end{aligned}
\end{equation}
where $\operatorname{Re}$ denotes real part operator and $C_0 = 20 \log_{10} e$.

Equation (\ref{Eq_RytovApp}) is the final form for our proposed x3DPRA.

\subsection{Background Scattering Subtraction}
The model (\ref{Eq_RytovApp}) is approximate only and cannot be expected to perform well in arbitrary environments. Its success is predicated on a number of assumptions. One significant advantage of the technique, however, is that it is linear in $\raisedchi_{\text{RI}}(\mathbf{r})$. This allows us to use a temporal subtraction technique to remove background scattering from the target object and significantly enhance the quality of the resulting reconstruction. It also provides a self-calibration procedure. 

Start with an initial time instant $t_0$, when the contrast profile is written as $\chi_{\text{RI}}^{t_0}$ and the measured total power as $P^{t_0}_{m_t}(\mathbf{r}_{m_r})$. Then let there be a change in the profile across time duration $\Delta t$ so that the contrast profile at time instant $t_0+\Delta t$ is $\chi_{\text{RI}}^{t_0+\Delta t}$ and the measured total power be $P_{m_t}^{t_0+\Delta t}(\mathbf{r}_{m_r})$. The change in power and profile can then be related by using (\ref{Eq_RytovApp}) as
\begin{equation}
	\label{Eq_RIdBTBS}
	\begin{aligned}
		\Delta  & P_{m_t}(\mathbf{r}_{m_r})  = P^{t_0+\Delta t}_{m_t}(\mathbf{r}_{m_r}) - P^{t_0}_{m_t}(\mathbf{r}_{m_r})  \\ & = C_0 \operatorname{Re}\left( \frac{k_0^2}{E^i_{m_t}(\mathbf{r}_{m_r}) h(\theta_{m_r,m_t})} \right.\cdot \\
		&\left. \int_{{\cal P}} g\left( \mathbf{r}_{m_r}, \mathbf{r} \right) \Delta\chi_{\text{RI}}\left( \mathbf{r} \right) E^i_{m_t} \left( \mathbf{r} \right) h(\theta_{{m_r},r}) d\mathbf{r}^2 \right)
	\end{aligned}
\end{equation}
where $\Delta P_{m_t}(\mathbf{r}_{m_r})$ (in dB) is the change in RSS in the duration $\Delta t$ and,
\begin{equation}
	\label{deltaNu}
	\begin{aligned}
		\Delta \chi_{\text{RI}} = \Delta \operatorname*{Re}(\chi_{\text{RI}}) + j \Delta \operatorname*{Im}(\chi_{\text{RI}}) = \chi_{\text{RI}}^{t_0+\Delta t} - \chi_{\text{RI}}^{t_0}
	\end{aligned}
\end{equation}

Equation (\ref{Eq_RIdBTBS}) provides reconstruction with background subtraction. This has several advantages, as it can image changes at a given DOI while removing strong multipath scattering from the background. 

Background subtraction is effective because it leverages the strong spatial sparsity of indoor environments. Scattering from background structures such as floors and ceilings is usually separated from the objects of interest by free space (air), separating the objects in nearly all directions. Therefore, a change in the object usually does not significantly affect the scattering caused by the background. This implies that subtracting the stationary background scattering can remove it from the reconstruction. This is one of the reasons why xPRA in 2D is very effective, as background selection allows it to focus on changes in the environment only. The spatial sparsity also implies that we can efficiently solve the ill-posed imaging problem with minimal measurements by applying a sparsity constraint. 

Another major advantage of background subtraction is that no calibration is required, since all sensor calibration is subtracted out. For example, in \cref{Eq_RytovApp}, the incident power needs to be estimated. This is difficult to obtain and can typically only be performed in laboratory settings using an anechoic chamber. By leveraging background subtraction, this difficulty is removed since it is subtracted out. 

To apply the background subtraction technique, we acquire measurements when no objects of interest are present in the DOI to obtain the background-clutter wave field. Equating that to $t_0$ we then utilize (\ref{Eq_RIdBTBS}) by substituting $P_{m_t}^{t_0}(\mathbf{r}_{m_r}) = P^i_{m_t}(\mathbf{r}_{m_r})$, $P_{m_t}^{t_0+\Delta t_0}(\mathbf{r}_{m_r}) = P_{m_t}(\mathbf{r}_{m_r}), \chi_{\text{RI}}^{t_0}=0$ and $\chi_{\text{RI}}^{t+\Delta t_0} = \chi_{\text{RI}}$, where $P^i_{m_t}(\mathbf{r}_{m_r}), P_{m_t}(\mathbf{r}_{m_r}), \chi_{\text{RI}}$ are respectively the free-space incident power, total measured power and the contrast respectively. 

\section{Reconstruction Procedure}
\label{sec: RR}
Our inverse problem has been described in Section \ref{sub: Problem}, and along with 
(\ref{Eq_RIdBTBS})- (\ref{deltaNu}) we are now in a position to consider reconstructing $\alpha(\mathbf{r}) \in \cal{D}$. In this work, we perform reconstruction using four steps, which we describe as discretization, sparsity, optimization, and regularization in the following subsections.

\subsection{Discretization}
We discretize the DOI to obtain a numerical formulation for our approach. The 3D DOI is discretized into $N = n_x \times n_y \times n_z$ voxels, with the size of each  $\Delta v = \Delta d_x \times \Delta d_y \times \Delta d_z$. For the $n$th voxel, the center location is specified $\mathbf{r}_n$. For link $l$, we define voxel distances as
\begin{equation}
	\label{eq: distances}
	r_{m_t,n} = |\mathbf{r}_{m_t} -\mathbf{r}_{n}|; \hspace{.5cm} r_{m_r,n} = |\mathbf{r}_{m_r} -\mathbf{r}_{n}|
\end{equation}
where $ r_{m_t,n}$ and $r_{m_r,n}$ represent the distances between the transmitter, receiver to the $n$th voxel, respectively with angles  $\theta_{m_r,n}$ etc defined similarly.

For link $l$ formed by $m_t$ and $m_r$, let $\Delta y_l = \Delta P_{m_t}(\mathbf{r}_{m_r})$, rewriting (\ref{Eq_RIdBTBS}) with discretization we arrive at

\begin{equation}
\label{eq: Delta_yl}
    \Delta y_l = C_0 \cdot \mathrm{Re} \left(k_0^2 \sum_{\forall n} \psi_{m_t}^{(m_r, n)}  \Delta \chi_{\text{RI}}(\mathbf{r}_n) \right)
\end{equation}
where the kernel $\psi_{m_t}^{(m_r, n)}$ is defined as
\begin{equation}
	\psi_{m_t}^{(m_r,  n)}=\frac{ h\left(\theta_{m_r, n}\right) g\left(\mathbf{r}_{m_r},\mathbf{r}_n\right) E_{m_t}^i\left(\mathbf{r}_n\right) \Delta v}{h\left(\theta_{m_r,  m_t}\right) E_{m_t}^i(\mathbf{r}_{m_r})} 
\end{equation}

We can then define our inverse problem as estimating $\Delta \operatorname*{Re}(\chi_{\text{RI}})$ and  $\Delta \operatorname*{Im}(\chi_{\text{RI}})$ given the measurements $\Delta y_l$. Note that only $\Delta \operatorname*{Im}(\chi_{\text{RI}})$ is our target variable since we can obtain the attenuation parameter change from it directly using 
\begin{equation}
\Delta \alpha\left(\mathbf{r}_n\right)=-k_0 \Delta \operatorname*{Im}(\chi_{\text{RI}}\left(\mathbf{r}_n\right))
\end{equation}
We include the negative on the right-hand side for convenience, as we shall see later. It should also be noted that even if we used $\Delta \operatorname*{Re}(\chi_{\text{RI}})$, its reconstruction is distorted due to the presence of refraction as explained in Section III in~\cite{Amar_xPRA}. 

We can write our linear model for $M$ links in compact matrix form as
\begin{equation}
\label{eq: real_y}
    \mathbf{y} = \mathrm{Re} (\mathbf{G} \Delta\boldsymbol{\chi})
\end{equation}
where the power measurements for the $M$ links are written as $\mathbf{y} = [\Delta y_1, \Delta y_2, \ldots,\Delta y_M] \in \mathbb{R}^{M\times 1}$ and $\Delta\boldsymbol{\chi} = [\Delta\chi_{\text{RI}}(\mathbf{r}_1),\Delta\chi_{\text{RI}}(\mathbf{r}_2),\ldots,\Delta\chi_{\text{RI}}(\mathbf{r}_N)]^T \in \mathbb{C}^{N\times 1}$ is the contrast vector for $N$ voxels. The entry $G_{l,n}$ of complex weight matrix $\mathbf{G} \in \mathbb{C}^{M\times N}$ for the $l$th link and $n$th voxel is

\begin{equation}
        G_{l,n} = C_0k_0^2 \psi_{m_t}^{(m_r,n)}
\end{equation}

Expanding~\cref{eq: real_y} further, we obtain

   \begin{equation}
\mathbf{y}=-\underbrace{\operatorname{Im(\mathbf{G})\operatorname{Im}(\Delta \boldsymbol{\chi})}}_{\begin{array}{c}
\text { Change in } \\
\text { attenuation loss }
\end{array}}+\underbrace{\operatorname{Re(\mathbf{G})\operatorname{Re}(\Delta \boldsymbol{\chi})}}_{\begin{array}{c}
\text { Change in } \\
\text { scattering loss }
\end{array}}+\mathbf{n}'
\end{equation}
and by writing the term $\operatorname{Re}(\mathbf{G})\operatorname{Re}(\Delta \boldsymbol{\chi})$ as being part of the system noise, $\mathbf{n}$, as performed in xPRA (2D) work~\cite{RTI,Amar_xPRA,Amar_xRTI} we obtain the matrix equation
\begin{equation}
\label{eq: xPRA_pre}
    \mathbf{y} = -\operatorname{Im(\mathbf{G})\operatorname{Im}(\Delta \boldsymbol{\chi)}} + \mathbf{n}
\end{equation}


Finally, writing in terms of attenuation parameter and defining the attenuation change vector as $\Delta \boldsymbol{\alpha} = [\Delta \alpha_1,\Delta\alpha_2,\ldots,\Delta\alpha_N]^T \in \mathbb{R}^{N \times 1}$, we can rewrite~\cref{eq: xPRA_pre} as a system model for the attenuation vector as
\begin{equation}
\label{eq: xPRA}
    \mathbf{y} = \mathbf{W}_{\text{xPRA}}\Delta\boldsymbol{\alpha} + \mathbf{n}
\end{equation}
where x3DPRA weight matrix $\mathbf{W}_{\text{xPRA}} = \operatorname{Im}(\mathbf{G})/k_0$. 

\subsection{Sparse Matrix Formulation}
The matrix $\mathbf{W}_{\text{xPRA}} \in \mathbb{R}^{M \times N}$ quickly becomes large in dimension due to the 3D nature of the problem. Therefore, we also draw on our previous findings~\cite{Amar_xRTI} that we can make the matrix sparse without losing significant information. This is achieved by multiplying the matrix by a mask that removes insignificant terms. This mask is originally derived from wave propagation theory (using only the first Fresnel zone) and is used to reconcile xPRA with RTI to obtain xRTI~\cite{Amar_xRTI}.

The mask is given by an ellipse-selection matrix $\mathbf{S} \in \mathbb{R}^{M \times N}$, having elements $S_{l,n}$ for the $l$th link and $n$th voxel. 
	\begin{equation}
		\label{eq: ellipse}
		S_{l,n } = \mathbb{I}\left[r_{m_t,n} + r_{m_r,n} <{r_{m_t,m_r} + \Delta d} \right]
	\end{equation}
where $\mathbb{I}[\cdot]$ is the indicator function, and $\Delta d$ is the width parameter to control the ellipse size such that the ellipsoid captures the first Fresnel zone of link $l$.

The weight and selection matrices are then multiplied together in an element-wise fashion $\mathbf{W} = \mathbf{W}_{\text{xPRA}} \odot \mathbf{S}$. In doing so, for link $l$, we keep only the values of voxels near the LOS and assign $0$ to the others. Thus, $\mathbf{W}$ becomes a sparse matrix and our sparse matrix formulations of~\cref{eq: xPRA} becomes
\begin{equation}
\label{eq: re_xPRA}
     \mathbf{y} = \mathbf{W}\Delta\boldsymbol{\alpha} + \mathbf{n}
\end{equation}

\subsection{Optimization and Regularization}
\label{subsec: Op}

For inverse scattering problems, the number of measurements $M$ is typically much smaller than the number of unknowns or voxels $N$ ($M \ll N$) so that the problem \cref{eq: re_xPRA} is ill-posed. To leverage the underlying spatial sparsity, we formulate a least square optimization \cite{Opt} with regularization as 
\begin{equation}
\label{eq: opt}
    \Delta\boldsymbol{{\hat \alpha}}=\underset{\Delta\boldsymbol{\alpha}}{\operatorname{argmin~}} \frac{1}{2}\|\mathbf{y}-\mathbf{W} \Delta\boldsymbol{\alpha}\|_2^2+\gamma T(\Delta\boldsymbol{\alpha})
\end{equation}
where $\gamma$ is the regularization weight parameter and  $T(\mathbf{\Delta\boldsymbol{\alpha}})$ is Total Variation (TV) regularization term.

We utilize a recent 3D regularization approach, which we refer to as TVReg~\cite{3DTV}, and it is written as
\begin{equation}
    \label{eq: 3DTV}
    T_{\text{3D}}(\Delta\boldsymbol{\alpha})=\sum_{n=1}^N \Phi_\tau\left(D_n \Delta\boldsymbol{\alpha}\right)
\end{equation}
where $D_n \in \mathbb{R}^{3 \times N}$ is the discrete difference operator for the differences between voxel $n$ and its neighbors. $\Phi_\tau$ is the Huber function, which is widely used in optimization problems to make the function smoother. Specifically, $D_n \mathbf{x}$ is
\begin{equation}
    D_n \Delta\boldsymbol{\alpha}= \left[\begin{array}{c}\nabla_x \Delta \alpha_n \\ \nabla_y \Delta \alpha_n \\ \nabla_z \Delta \alpha_n\end{array}\right]
\end{equation}
where $\nabla_x \Delta \alpha_n = \Delta \alpha_{n+1}-\Delta \alpha_n$ is the difference along the x dimension and similarly for, $\Delta_y \Delta \alpha_n$, and $\nabla_z \Delta \alpha_n$.

The Huber function $\Phi_\tau(\mathbf{z})$ is
\begin{equation}
    \label{eq: Huber}
    \Phi_\tau(\mathbf{z})= \begin{cases}\|\mathbf{z}\|_2-\frac{1}{2} \tau, & \text { if }\|\mathbf{z}\|_2 \geq \tau, \\ \frac{1}{2 \tau}\|\mathbf{z}\|_2^2, & \text { else. }\end{cases}
\end{equation}
where $\tau$ is a constant threshold. The function applies the Euclidean norm for large values $\|\mathbf{z}\|_2 \geq \tau$, while calculating a quadratic value for smaller $\mathbf{z}$.

For model verification, we also wish to consider cross-sections, resulting in a 2D formulation of (\ref{eq: opt}). The cross sections have $N'$ voxels in a plane where $N' < N$. For this we use TVAL3~\cite{R_2DTV} with regularization term 
\begin{equation}
    \label{eq: 2D TV}
    T_{\text{2D}}(\Delta\boldsymbol{\alpha}) = \sum_{n=1}^{N'} ||D_{n} \Delta\boldsymbol{\alpha}||_2
\end{equation}
where $||\cdot||_2$ refers to norm-2, and $D_n \in \mathbb{R}^{2\times N'}$ is the discrete difference operator in 2D. Similarly, $D_n \Delta\boldsymbol{\alpha}$ is the difference operator in $x$ and $y$ dimensions only. 

\section{Simulation Results}
\label{sec: R}
This section provides simulation results for x3DPRA as defined in~\cref{eq: opt}. To provide a comprehensive analysis, the results are divided into two parts: model verification and 3D reconstruction. Before progressing to full 3D scenarios, we first verify x3DPRA in 2D using ${\cal M}_{2D}$. In the 3D simulations, we generate ${\cal M}_{3D}$ measurement data by using CST Studio Suite.

\subsection{Parameter Setting}
To configure the system to resemble a WiFi scenario, we set the frequency to 2.4 GHz ($\lambda_0=0.125$ m). The CST bounding box used is $0.96 \times 0.96 \times 0.86$ $\text{m}^3$. The 3D DOI fits within this box and is restricted to $0.9\times 0.9 \times 0.3$ $\text{m}^3$, which corresponds to $7.2 \times 7.2 \times 2.4$ $\lambda_0^3$. The DOI is limited by the need to gather simulated measurement data rather than by the reconstruction approach. Vertically polarized dipole antennas are placed around the boundary of the DOI, and when connected to transceivers, form nodes.  

\begin{figure}[htbp]
	\centering
	\includegraphics[width=1\linewidth]{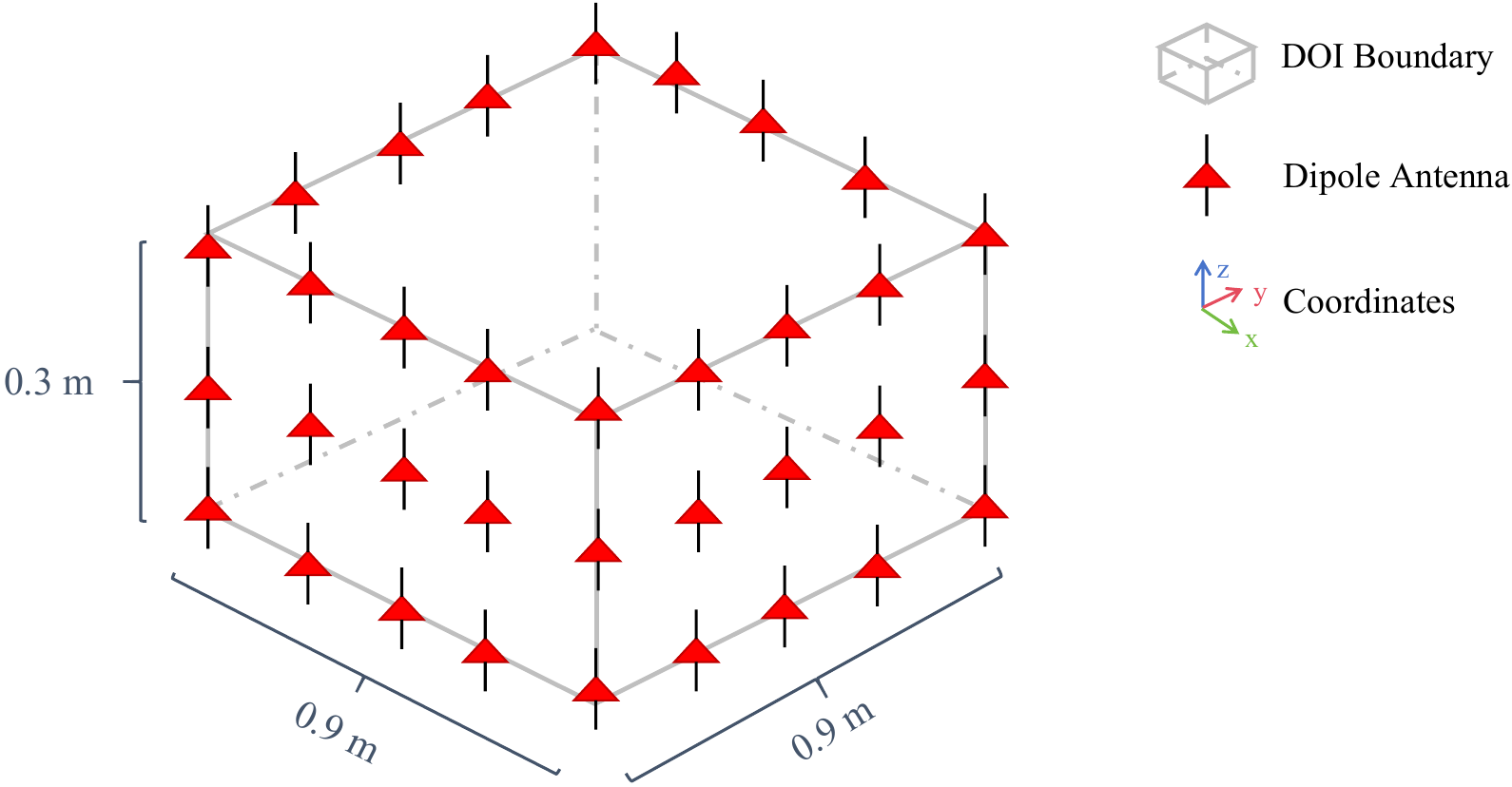}
	\caption{The 3D DOI has dimensions of $0.9\times 0.9 \times 0.3$ $\text{m}^3$ and sits within a CST bounding box of  $0.96 \times 0.96 \times 0.86$ $\text{m}^3$. A total of $L = 48$ transceivers are equally distributed around the boundary at each of three heights: $z = 0, \pm0.15$ m. For clarity, only the transceivers positioned on the front face of DOI are depicted in this figure.} 
	\label{fig: TRx}
\end{figure}

For clarity, we express the 3D DOI in rectangular coordinates. The center of the DOI is defined as coordinate $(0,0,0)$, with $x$ and $y$ dimension ranging from $[-0.45, 0.45]$ m, and the z dimension within $[-0.15, 0.15]$ m.  Fig~\ref{fig: TRx} presents an empty DOI with the maximum number of $L = 48$ dipole antennas. Each dipole antenna has a length of $6$ cm and is placed at three different heights, $z = 0, \pm0.15$ m. At each of these three heights, $16$ antennas are evenly distributed around the boundary. For all simulation settings, the distance between the nearest neighboring antennas is greater than one wavelength to minimize antenna coupling.

\subsection{Measurement Configurations}
\subsubsection{2D}

\begin{figure}[htbp]
	\begin{subfigure}[b]{\linewidth}
		\centering
		\includegraphics[width=\textwidth]{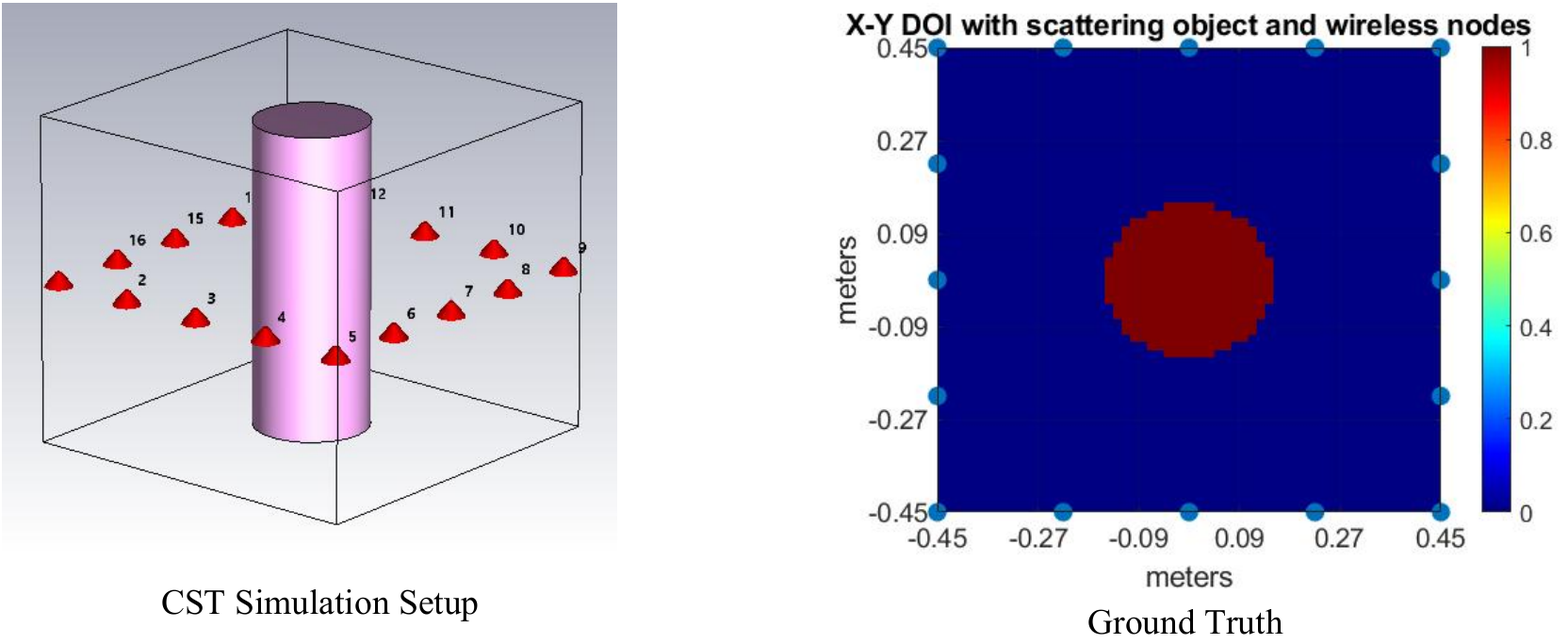}
		\caption{The ground truth cross-section is a circle.}
		\label{fig: VGT_Cir}
	\end{subfigure}
	\begin{subfigure}[b]{\linewidth}
    \centering
		\includegraphics[width=\textwidth]{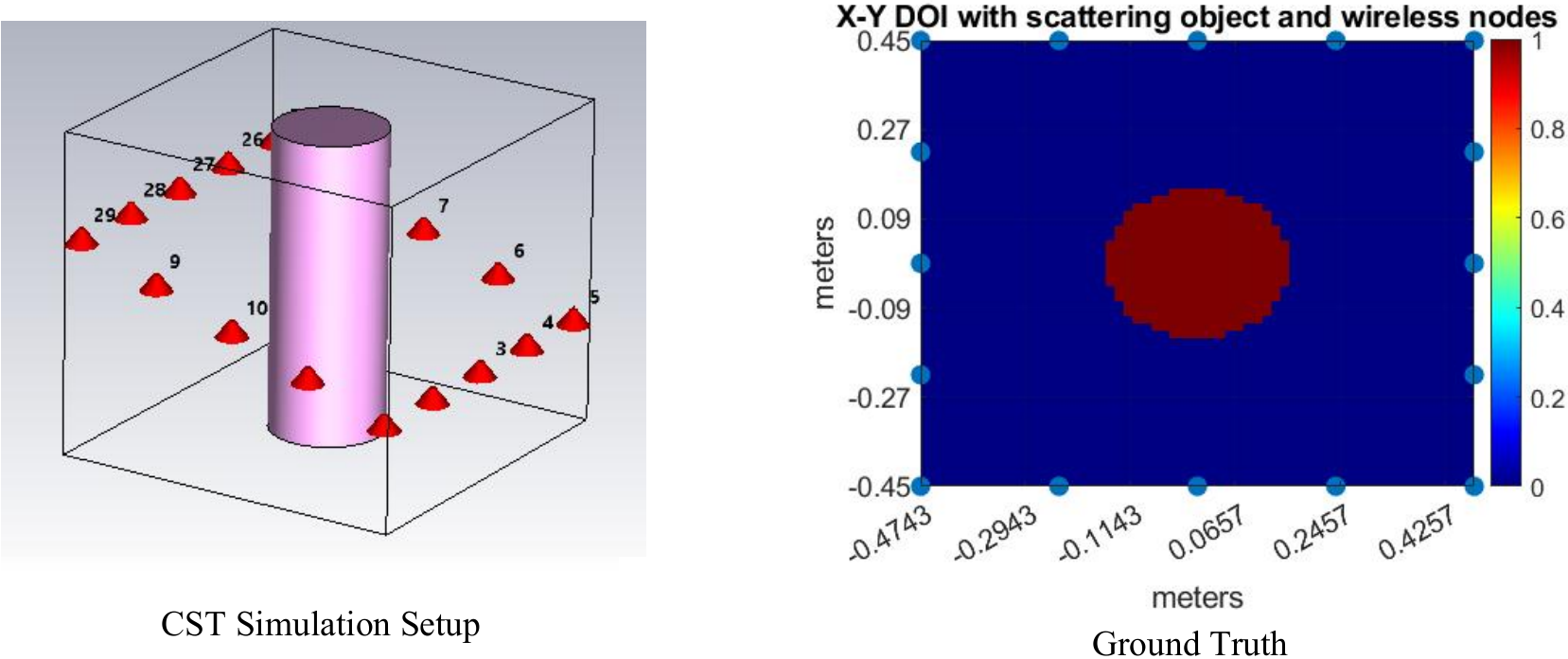}
		\caption{The inclined plane ground truth cross-section is  an 
			ellipse centered at $(0,0)$ with width $0.316$ m and length $0.3$ m.}
		\label{fig: VGT_Ell}
	\end{subfigure}
	\caption{16 Transceiver nodes are placed on a single center plane to form a 2D scenario. The circular cylinder has a cross-section of a circle
		centered at $(0,0)$ with diameter $0.3$ m.}
	\label{fig: VGT}
\end{figure}

To provide verification of our technique, we first consider 2D examples using 2D measurement sets ${\cal M}_{2D}$. Two example objects are provided, and these are shown in  Fig~\ref{fig: VGT}. The object used in both examples is the same, which is a circular cylinder with a diameter of $0.3$ m (or $2.4 \lambda_0$) and vertical height $0.8$ m (or $6.4 \lambda_0$).  Its center coincides with the DOI center. The object is designed to exceed the DOI height, so it better approximates a 2D system. The permittivity of the cylinder is $10 + 1j$, and the corresponding attenuation parameter is $\alpha = 15.8$.

In the first scenario, the antenna configuration is shown in Fig~\ref{fig: VGT_Cir}, where 16 transceivers are distributed equally around the boundary of DOI at the height of $z = 0$ m. Although the object is 3D, the scenario performs a pseudo-2D validation with a size of $0.9 \times 0.9$ $\text{m}^2$. We focus on reconstructing the circular shape, as shown in Fig~\ref{fig: VGT_Cir}. For model verification, we only aim to test the accuracy of location and shape reconstruction, so the values of the ground truth in the 2D results are all normalized. The attenuation values are estimated in the 3D reconstruction. 

In the second configuration, we use an inclined planar transceiver arrangement as shown in Fig~\ref{fig: VGT_Ell}. The inclined plane with size $0.95 \times 0.9$ $\text{m}^2$ is defined with its four corners: $(-0.45, -0.45, 0.15)$, $(-0.45, 0.45, 0.15)$, $(0.45, -0.45, -0.15)$, $(0.45, 0.45, -0.15)$. A total of $16$ nodes are equally spaced around the plane that intersects the cylinder as an ellipse. Fig~\ref{fig: VGT_Ell} presents the ground truth ellipse. 

We can initially verify x3DPRA's performance through the two setups in Fig~\ref{fig: VGT}. The reconstruction resolution for both setups is $N' = n_x \times n_y = 60\times 60$, with a total number of measurements $M = 16\times 15/2 = 120$. There is no resolution for the $z$ dimension because the reconstruction is limited to a single plane. 

\subsubsection{3D}
\begin{figure}[htbp]
    \begin{subfigure}[b]{\linewidth}
        \centering  \includegraphics[width=0.6\linewidth]{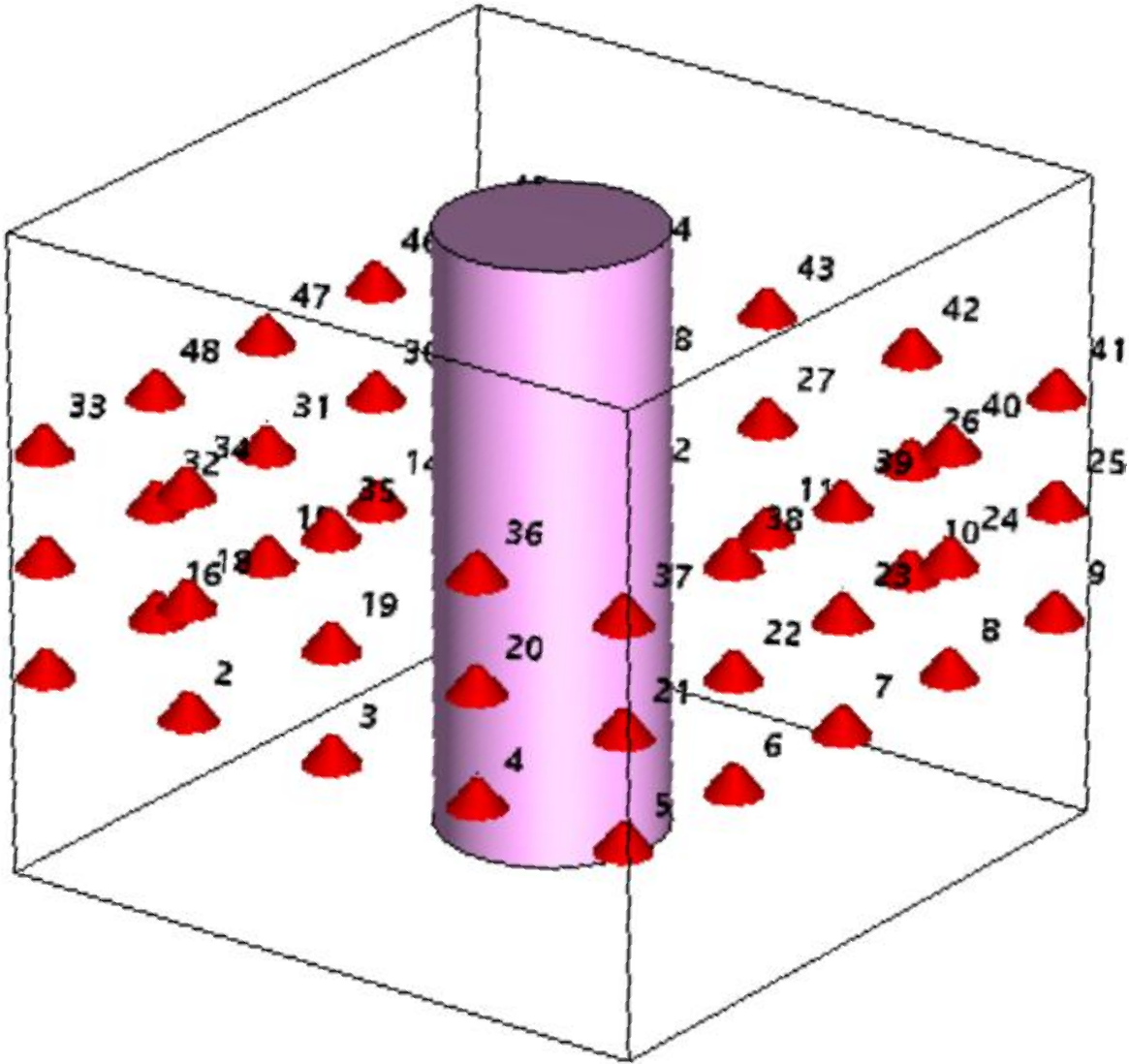}
         \caption{A $0.8$ m tall circular cylinder with $0.3$ m diameter  centered in the DOI.}
         \label{subfig: Cylin_GT}
    \end{subfigure}
    \vfill
    \begin{subfigure}[b]{\linewidth}
        \centering       \includegraphics[width=0.6\linewidth]{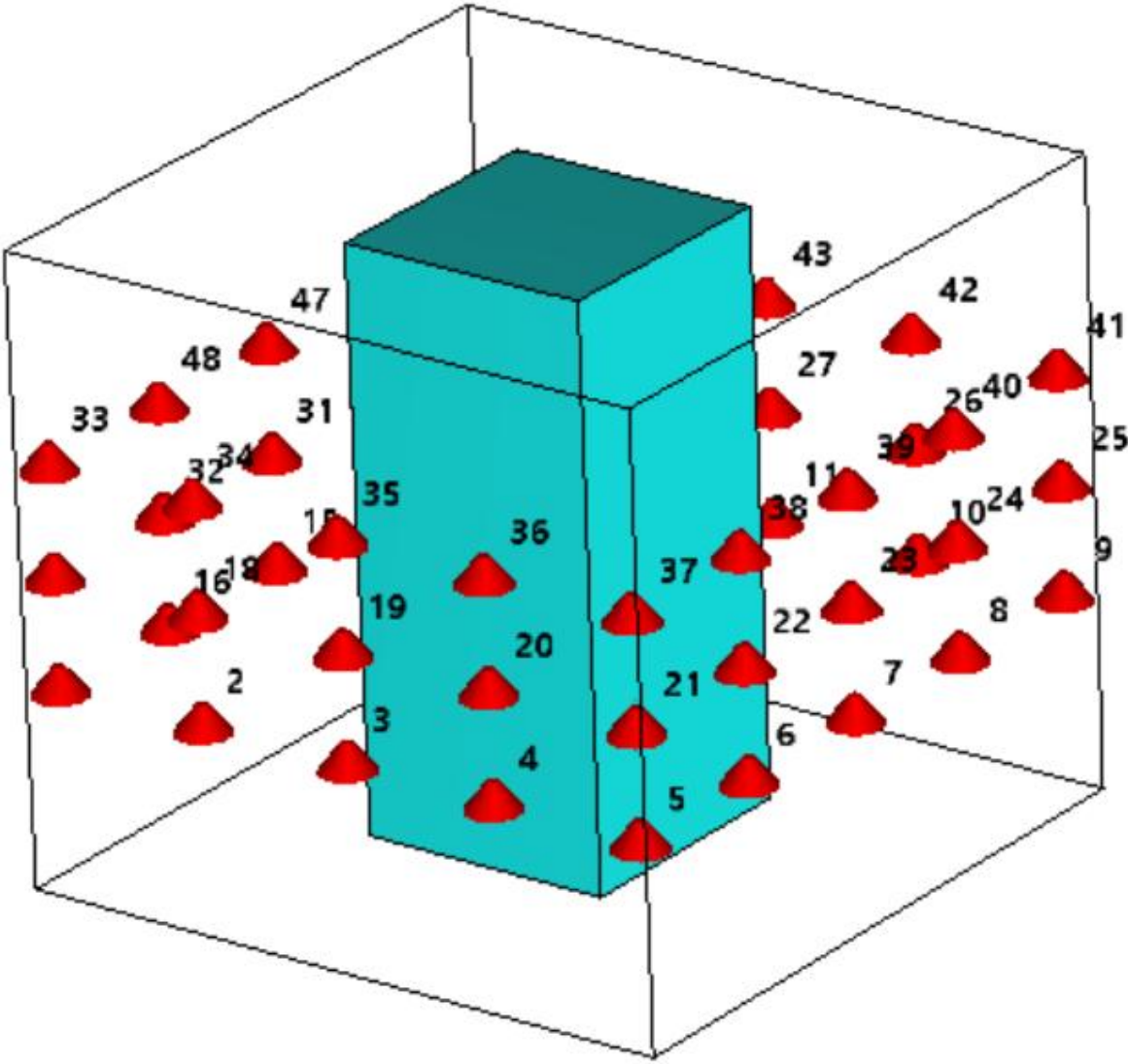}
         \caption{A $0.8$ m tall square cylinder with sides $0.36$ m length  centered in the DOI.}
         \label{subfig: Cuboid_GT}
    \end{subfigure}
    \vfill
    \begin{subfigure}[b]{\linewidth}
        \centering      \includegraphics[width=0.6\linewidth]{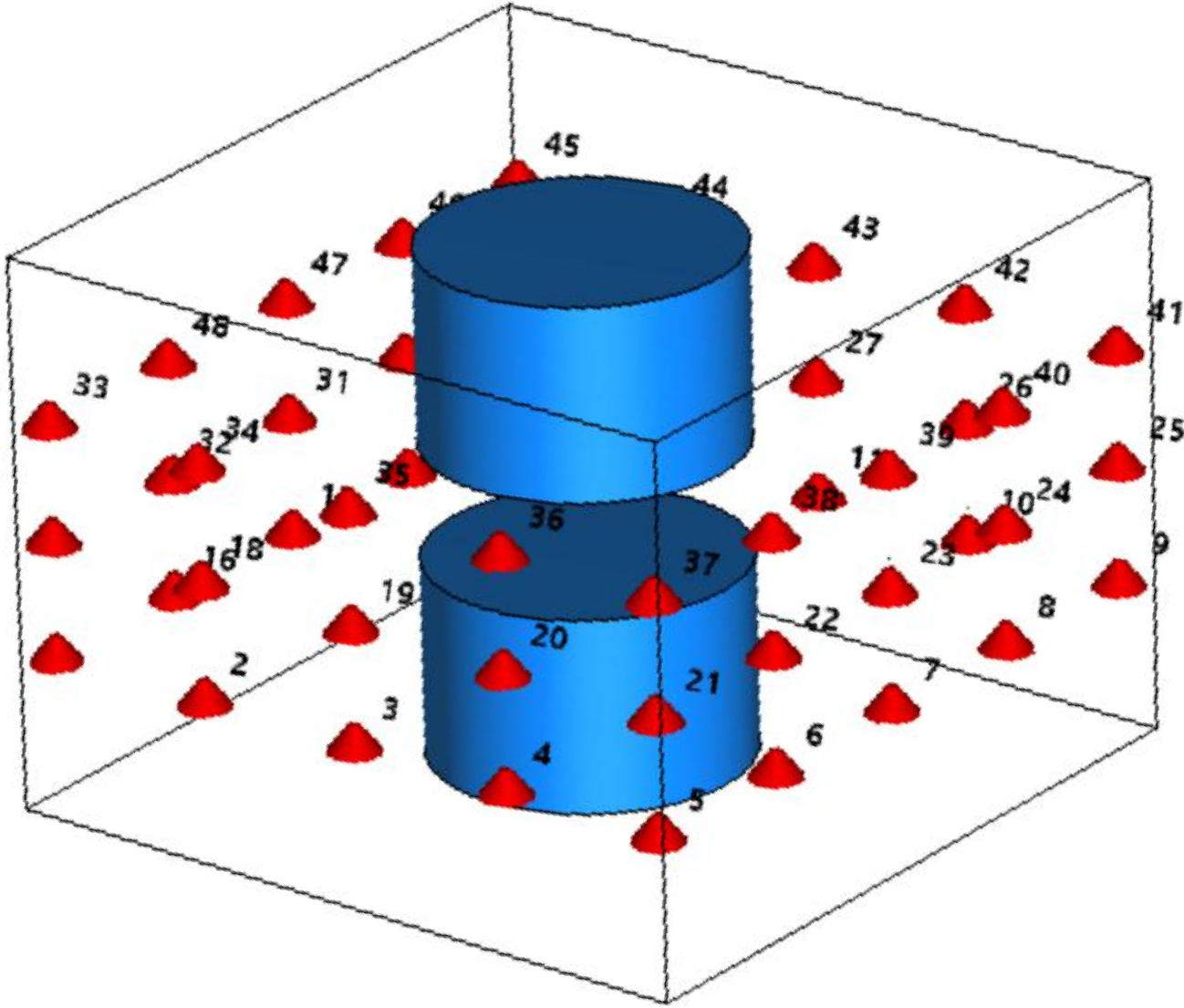}
         \caption{Two $0.25$ m tall circular cylinders with $0.4$ m diameter, are located at separate centers $(0, 0, \pm 0.2)$ m.}
         \label{subfig: 2CylindersV}
    \end{subfigure}
    \caption{The configurations used for the 3D configurations, where the red markers are the center positions of the antennas. The antennas are distributed equally around the boundary at heights $z = 0, \pm 0.15$ m. }
    \label{fig: GT}
\end{figure}
For 3D reconstruction, we have three scenarios as shown in Fig~\ref{fig: VGT}. The first is the same circular cylinder as shown in Fig~\ref{subfig: Cylin_GT}. The second is a square cylinder centered within the DOI with a length of $0.36$ m (or $2.9 \times \lambda_0$), height $0.8$ m (or $6.4 \times \lambda_0$), and is shown in Fig~\ref{subfig: Cuboid_GT}. The permittivity of the object is $8+0.8j$ with a corresponding attenuation parameter of $\alpha = 14.2$. The third object is shown in Fig~\ref{subfig: 2CylindersV}, which is composed of two cylinders with a diameter of $0.4$ m (or $3.2 \times \lambda_0$) and a height of $0.25$ m (or $2\times\lambda_0$). The center of the bottom cylinder is at $(0,0,-0.2)$ m, while the center of the top cylinder is at $(0,0,0.2)$ m. There is a $0.15$ m vertical space ($> \lambda_0$) between them to reduce reflection and scattering. The permittivity of this object is $15+1.5j$, and the attenuation parameter is $\alpha = 19.5$. The simulation arrangement of $48$ transceivers at heights $z = 0, \pm0.15$ m and ground truth for these objects in 3D reconstruction are demonstrated in Fig.~\ref{fig: GT}. 

To save CST simulation time, we use symmetry by mirroring the measurement data from $32$ transceivers at $z =- 0.15$ m and $z = 0$ m to generate a corresponding set of virtual measurements at $z = 0$ m and $z = 0.15$ m. Thus, there are $M = 2 \times \frac{32\times 31}{2} = 992$ measurements. 

All objects have a loss tangent of $\delta = 0.1$. The permittivity of the objects can be calculated through~\cref{eq: att} once we obtain the reconstructed attenuation parameter. For 3D reconstruction, the resolution is $n_x \times n_y \times n_z = 60\times 60 \times 15$.

\subsection{Results}
\subsubsection{2D Reconstructions}
\begin{figure}[htbp]
    \includegraphics[width=0.5\textwidth]{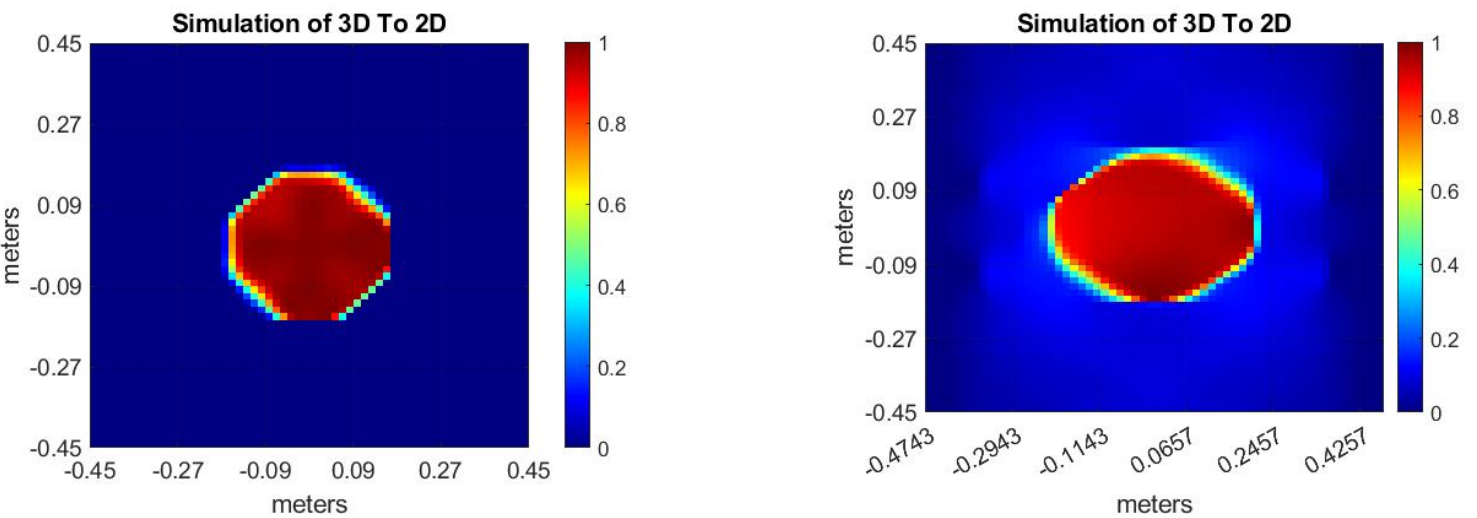}
    \caption{The result on the left corresponds to the circular ground truth in Fig~\ref{fig: VGT_Cir}, with PSNR $19.0$ dB; while the result on the right corresponds to the ellipse ground truth in Fig~\ref{fig: VGT_Ell}, with PSNR $11.5$ dB.}
    \label{fig: VR}
\end{figure}

The model verification results using TV regularization~\cref{eq: opt} with 2D TV solver TVAL3~\cref{eq: 2D TV}~\cite{R_2DTV} are shown in Fig~\ref{fig: VR}. All results are normalized during model verification for simplicity, and only the location and shape are indicated. The regularization parameter $\gamma$ in TVAL3~\cite{R_2DTV} is set as $\gamma = 2^p$ (for $p \in \mathbb{R}$). The results shown use the $\gamma$ that yields the highest reconstruction PSNR.

 Compared to Fig~\ref{fig: VGT_Cir}, the result on the left is consistent with the ground truth. For the result on the right the result also aligns with the ground truth in Fig~\ref{fig: VGT_Ell} for the inclined plane. 

These results verify the model's performance and accuracy, providing a solid foundation for testing in 3D scenarios. 

\subsubsection{3D Reconstructions}
\begin{figure}[htbp]
\begin{subfigure}[b]{\linewidth}
    \centering
        \includegraphics[width=\textwidth]{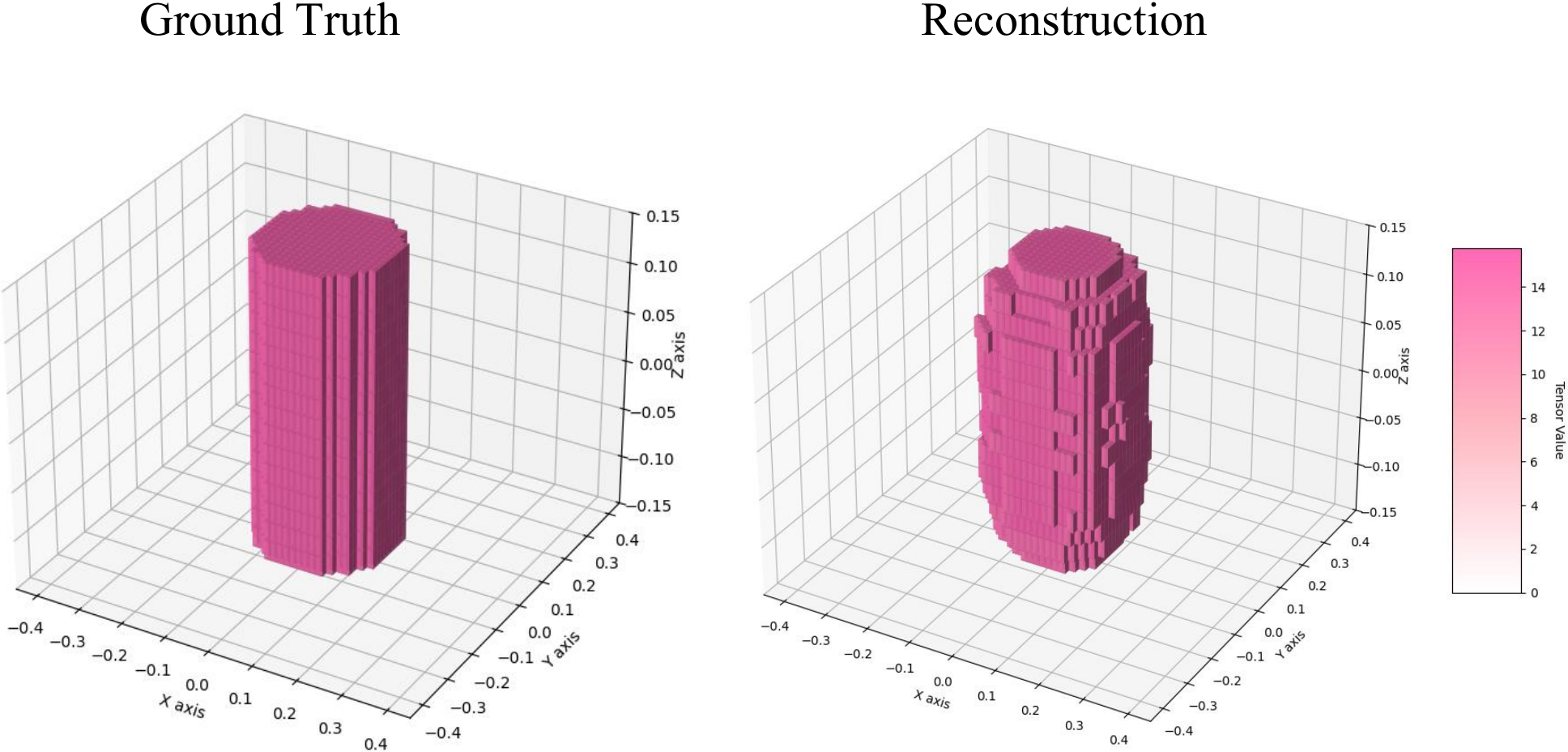}
        \caption{The result is for the circular cylinder ground truth in Fig~\ref{subfig: Cylin_GT}, which provides a PSNR value of $19.8$ dB}
    \label{subfig: R_Cylin}
    \end{subfigure}
    \begin{subfigure}[b]{\linewidth}
    \centering        \includegraphics[width=\textwidth]{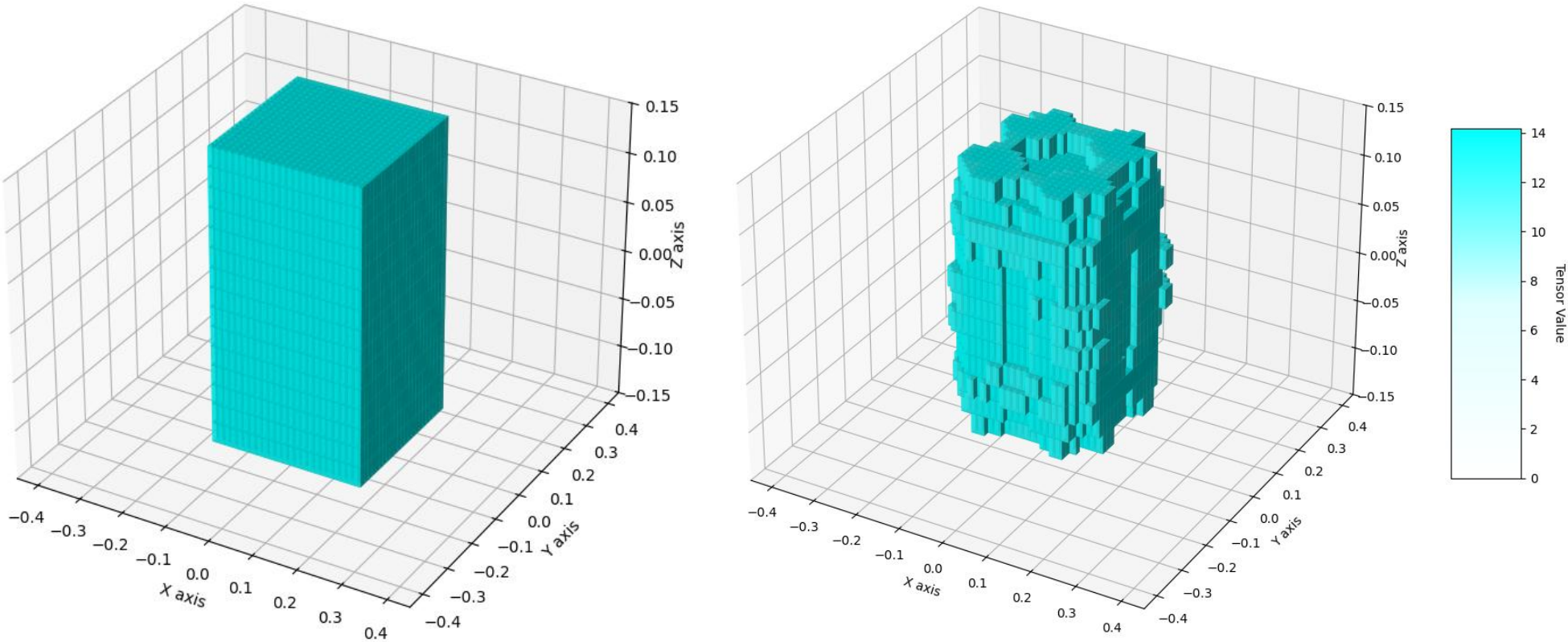}
        \caption{The result is for the square cylinder ground truth in Fig~\ref{subfig: Cuboid_GT}, which provides a PSNR value of $16.3$ dB}
    \label{subfig: R_Cuboid}
    \end{subfigure}
    \begin{subfigure}[b]{\linewidth}
    \centering
    \includegraphics[width=\textwidth]{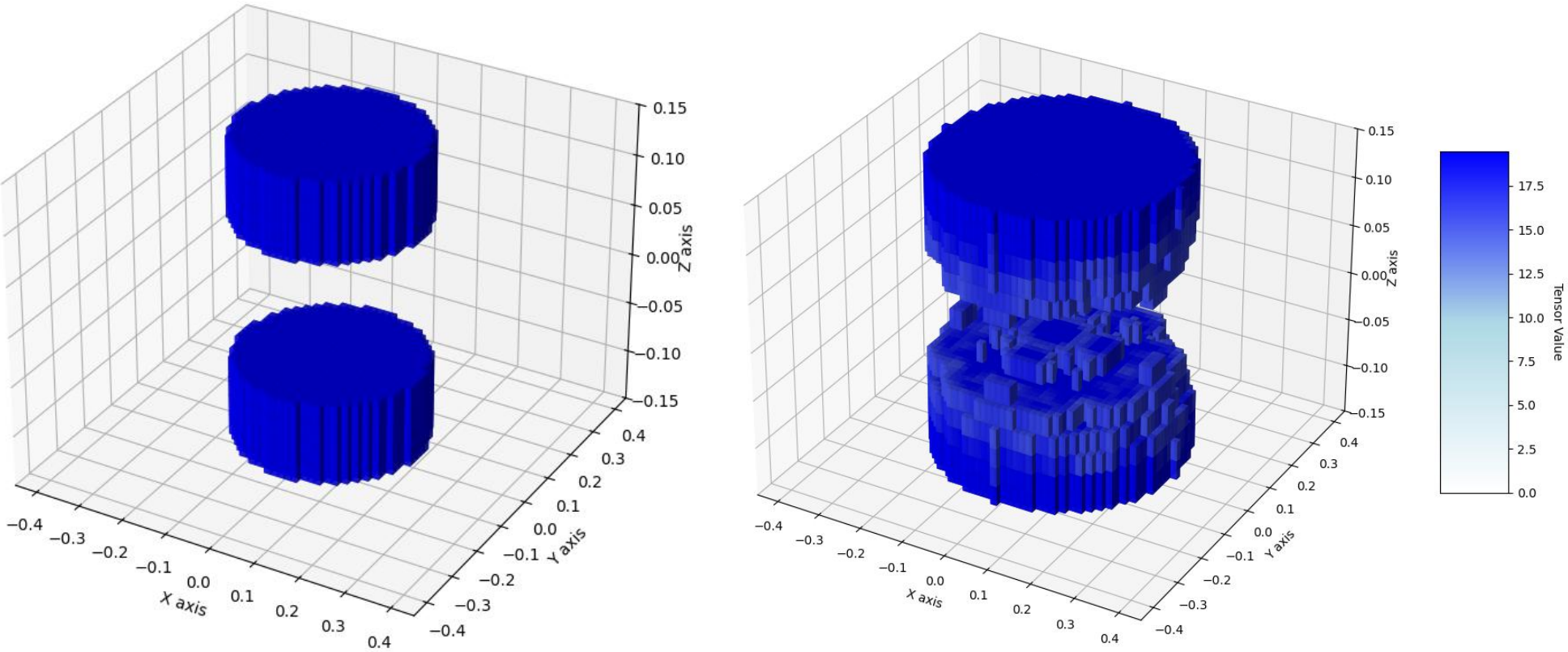}
        \caption{The result is for the two circular cylinder ground truth in Fig~\ref{subfig: 2CylindersV}, which provides a PSNR value of $10.0$ dB.}
        \label{subfig: R_2CylinV}
    \end{subfigure}
    \caption{The figures show the 3D reconstruction results and their corresponding ground truth. For all figures, the first column shows the ground truth, and the second shows the reconstruction results.}
    \label{fig: 3D_R}
\end{figure}
We obtain the 3D reconstruction results by solving~\cref{eq: opt} with the 3D TV solver TVReg~\cref{eq: 3DTV}~\cite{3DTV}. The regularization parameter $\gamma$ in TVReg~\cite{3DTV} and the threshold for the Huber function~\cref{eq: Huber} were fine-tuned to maximize the PSNR in reconstruction. Fig~\ref{fig: 3D_R} provides the reconstruction of the three objects; the first column is the ground truth, and the second column contains the 3D reconstruction results. Generally, the 3D reconstructions show good agreement with their corresponding ground truth in terms of shape, location, and attenuation values.

For the circular cylinder in Fig.~\ref{subfig: R_Cylin}, the results are generally well-aligned with the ground truth. Although the reconstruction volume is similar to the ground truth, its radius is smaller near the top ($z = 0.15$ m) and bottom ($z = -0.15$ m) planes. It is related to the LOS values and the TV property. The measured attenuation correlates with a link's interaction with the cylinders. Links confined to the $z = 0, \pm 0.15$ m plane show minimal loss. LOS paths passing through different cross sections exhibit greater attenuation as they traverse a longer length of the object. There is a direct relationship between attenuation loss and reconstruction volume: higher attenuation loss yields larger volumes. At the same time, TV regularization aims to minimize the differences between neighboring voxels. In the $x-z$ direction, TV aims to make each layer similar in size to its adjacent layers, resulting in a decrease in size toward the top and bottom compared to the middle. For the attenuation parameter reconstruction, the results yield the ground truth value of $\alpha=15.8$ throughout most of the volume.

For the square cylinder in Fig.~\ref{subfig: R_Cuboid}, we can observe that the attenuation parameter reconstruction yields the expected value $\alpha = 14.2$ throughout most of the imaged volume. While the shape of the results is similar to the ground truth, the overall performance exhibits some mismatches. It is mainly due to the sharp corners of the square cylinder and the reflections they cause. At the corners, the LOS path passing through these angles experiences attenuation similar to that of the nearby LOS, leading the model to reconstruct a rounder shape. The TV term also rounds the cuboid's sharp corners and shrinks its flat faces to avoid gradient penalties, producing a slightly smaller reconstructed width. Similar inaccuracies in reconstructing sharp geometric features have been observed in simpler 2D cases, where even accurate 2D models such as xPRA (2D) do not always yield precise reconstructions for square objects~\cite{Amar_xPRA, Amar_xRTI}. 

For two circular cylinders separated vertically in Fig~\ref{subfig: R_2CylinV}, the reconstruction is generally consistent with the ground truth by maintaining the cylinders' shape and attenuation value around $19.5$. However, there are two minor discrepancies. First, the reconstructed shape is slightly wider than the ground truth. This is because the TV term (the second term in~\cref{eq: opt}) encourages the solution to minimize the differences between voxels near the object's boundaries. When the data term (the first term in~\cref{eq: opt}) requires two disconnected structures, the optimizer compensates for the high penalty associated with multiple boundaries by increasing the radius of each cylinder. Second, the reconstruction exhibits uneven shapes and lower attenuation parameter values near the boundaries of objects at heights $z = \pm 0.075$ m. This behavior stems from the same factors affecting the LOS values and TV regularization in the single-cylinder case. The LOS paths from $z = 0$ m to $z = \pm 0.15$m pass through the objects' boundaries at $z = \pm 0.075$ m and experience significant attenuation. In contrast, LOS paths at $z = 0$ m travel only through air and thus experience nearly no loss. Consequently, the reconstructed attenuation is near zero for voxels on the plane $z = 0$ m but substantially higher for those on the planes $z = \pm 0.075$ m. However, the TV term penalizes abrupt differences between neighboring voxels, encouraging smooth transitions. To satisfy this constraint, the optimizer produces an uneven shape and lowers the reconstructed values in the vertical ranges $z \in (-0.075, 0) \cup (0,0.075)$ m as a trade-off. One possible solution is to combine x3DPRA with machine learning to minimize loss using iterative algorithms.

Overall, x3DPRA successfully reconstructs the shape, location, and attenuation parameters of three different objects. To the best of our knowledge, x3DPRA is the first 3D imaging model that uses phaseless measurements to provide good 3D volume reconstruction of objects with different permittivities of these values and object sizes. 
\section{Conclusion}
\label{sec: C}
We have developed a novel framework for device-free 3D radio tomographic imaging by extending xPRA from 2D to 3D. We have derived the x3DPRA model and implemented a corresponding optimization algorithm to reconstruct 3D attenuation maps. Our results demonstrate good localization and volumetric shape reconstructions under different settings. This represents progress in the field of RTI, as x3DPRA can provide height information, which is crucial for real-world applications. Moreover, we demonstrate the model's potential for material identification by estimating attenuation parameters for various objects. The proposed x3DPRA framework directly addresses a gap in wireless sensing: existing 3D propagation-based methods either provide low-resolution images or require multiple frequencies. This work proposes a sensing modality using electromagnetic inverse scattering for ISAC.

Future work will focus on experimental validation for x3DPRA. After validation, we consider several combinations with x3DPRA, including radar, beamforming, and machine learning. Specifically, integrating x3DPRA with radar technology~\cite{C_radar} could potentially further improve reconstruction by utilizing both propagation and reflection data in 3D. By adding beamforming~\cite{C_beam}, we can focus on a specific target in 3D DOI. With machine learning~\cite{C_machine}, x3DPRA enables real-time reconstruction for real-world applications. Artificial Intelligence (AI) techniques are also a prime avenue for further enhancing reconstruction quality through physics-aware formulations.

\bibliography{references}

\end{document}